\documentclass[11pt,twoside]{article}
\usepackage{./asp2014}

\aspSuppressVolSlug
\resetcounters

\begin{document}

\title{Solar Simulations for the Atacama Large Millimeter Observatory Network} 

\author{Sven Wedemeyer,$^{1,4}$ Tim Bastian,$^2$ Roman Braj\v sa,$^{3,4}$ Miroslav Barta,$^4$ Masumi Shimojo$^5$
\affil{$^1$Institute of Theoretical Astrophysics, University of Oslo, Norway;  \email{sven.wedemeyer@astro.uio.no}}
\affil{$^2$National Radio Astronomy Observatory (NRAO), Charlottesville, VA, USA} 
\affil{$^3$Hvar Observatory, Faculty of Geodesy, University of Zagreb, Croatia}
\affil{$^4$European ALMA Regional Center - Czech node, Astronomical Institute of Academy of Sciences, Ondrejov, Czech Republic}
\affil{$^5$National Astronomical Observatory of Japan (NAOJ),  Japan} 
}

% This section is for ADS Processing.  There must be one line per author.
\paperauthor{Sven~Wedemeyer}{sven.wedemeyer@astro.uio.no}{http://orcid.org/0000-0002-5006-7540}{University of Oslo}{Institute of Theoretical Astrophysics}{Oslo}{}{0315}{Norway}
%\paperauthor{Sample~Author2}{Author2Email@email.edu}{ORCID_Or_Blank}{Author2 Institution}{Author2 Department}{City}{State/Province}{Postal Code}{Country}
%\paperauthor{Sample~Author3}{Author3Email@email.edu}{ORCID_Or_Blank}{Author3 Institution}{Author3 Department}{City}{State/Province}{Postal Code}{Country}

\begin{abstract}
The Atacama Large Millimeter/submillimeter Array (ALMA) will be a valuable tool for observing the chromosphere of our Sun at (sub-)millimeter wavelengths at high spatial, temporal and spectral resolution and as such has great potential to address long-standing scientific questions in solar physics.
In order to make the best use of this scientific opportunity, the  Solar Simulations for the Atacama Large Millimeter Observatory Network has been initiated. 
A key goal of this international collaboration is to support the preparation and interpretation of future observations of the Sun with ALMA.
\end{abstract}

%================================================================================
%================================================================================
%================================================================================
\section{ALMA observations of the Sun} 
%================================================================================
%================================================================================
%================================================================================

Owing to its impressive capabilities, observations of the Sun with the Atacama Large Millimeter/submillimeter Array (ALMA) promise many ground-breaking scientific results. 
Most of the radiation at sub-/millimeter wavelengths is emitted in the Sun's chromosphere, which is a region sandwiched between the photosphere below and the transition region/corona above. 
The chromosphere plays a key role in the transport of mass and energy in the solar atmosphere but, owing to its complex structure and dynamics, many aspects remain elusive. 
ALMA now acts as a new tool, which will give us important clues for a number of still open but essential questions 
\citep{bastian02,2011SoPh..268..165K}. 
Potential scientific applications include the dynamics, thermal and magnetic structure and energy transport in the ``quiet'' solar chromosphere, in active regions and in sunspots, spicules, prominences and filaments, and flares with implications for the coronal heating problem, space weather and the atmospheres of other stars.
See \cite{ssalmon_ssrv15} for a comprehensive review over solar science cases for ALMA.

ALMA solar observing modes, which are currently being defined and tested, have different requirements from normal observing. 
For instance, the time scales on which the solar chromosphere evolves can be as short as few ten seconds so that integration over long periods or exploiting the EarthÕs rotation for increasing the u-v coverage are not an option for many science cases.
Moreover, interferometric imaging of an extended area source like the Sun is challenging in itself but yields unprecedented measurements of the thermal and magnetic structure of the chromosphere. 
In this respect, numerical simulations of the solar chromosphere \citep[e.g.,][]{2007A&A...471..977W,2015A&A...575A..15L} play an important role for the planning, optimizing and interpretation of observations with ALMA. 
Synthetic brightness temperature maps, which are calculated based on such models, can be used to simulate what ALMA would observe. 
Different instrumental set-ups can be tested and adjusted to the scientific requirements, finding the optimal set-up for individual scientific applications. 
In addition, simulations can demonstrate what could be possibly observed with ALMA and which scientific problems could therefore be addressed in the future.

%================================================================================
%================================================================================
%================================================================================
\section{SSALMON - An international network} 
%================================================================================
%================================================================================
%================================================================================

In support of the preparation of regular ALMA observations of the Sun, the 
\textit{Solar Simulations for the Atacama Large Millimeter Observatory Network}\footnote{More information and registration at \url{http://ssalmon.uio.no}.} (SSALMON) 
was initiated on September~1st, 2014  \citep{2015arXiv150205601W}. 
As of February 2015, more than 50~scientists from 15~countries are participating. 
The scientific network is organised in connection with two international development studies, namely  
(1)~``\textit{Advanced Solar Observing Techniques}'' (a project within the North American Study Plan for Development Upgrades of the ALMA; PI: T. Bastian, NRAO, USA), and 
(2)~``\textit{Solar Research with ALMA}'' (a project carried out at the Czech node of European ALMA Regional Center at Ondrejov in the frame of the ESO program ``\textit{Enhancement of ALMA Capabilities/EoC}''; PI: R. Brajsa, Hvar Observatory, Croatia).

The key goals of the SSALMONetwork  include 
(1)~raising awareness of science opportunities with ALMA within the solar physics community,  
(2)~increasing the visibility of solar science within the general ALMA community, and 
(3)~constraining ALMA observing modes through modeling efforts. 
The network activities thus focus on all related simulation and modelling aspects ranging from calculating numerical models of the solar atmosphere, producing synthetic brightness temperature maps, applying instrumental effects, comparisons with real ALMA observations of the Sun to developing optimized observation strategies.
For this purpose, a number  of  expert teams is currently being established, which will work on specific aspects. 
% like, e.g., radiative transfer and brightness temperature synthesis, simulating instrumental effects for ALMA (incl. interferometric imaging),  spectral lines in the millimeter range as new diagnostic tools,  and magnetic field measurements. 
%
The results will be published in a forthcoming paper series.

%================================================================================
%================================================================================
%================================================================================
%\bibliography{swb}  % For BibTex
% For non-BibTex:

\end{document}